\documentclass{article}
\usepackage{latexsym}
\usepackage{graphicx}
\title{Dynamic Relationship between Diversity and Plasticity of Cell types
in Multi-cellular State}
\author{Hiroaki Takagi, Kunihiko Kaneko\\
{\it Department of Pure and Applied Sciences}\\
{\it University of Tokyo, Komaba, Meguro-ku, Tokyo 153, JAPAN}}
\begin{document}
\maketitle
\begin{abstract}
Dynamics maintaining diversity of cell types in a multi-cellular system
are studied in relationship with the plasticity of cellular states.
First, we introduce a new theoretical framework, reaction-diffusion system 
on `chemical species space' to model intra-cellular chemical reaction
dynamics. Then, by considering the cell division and death of such
cells, developmental process from a single cell is studied.
Cell differentiation process is found to occur through instability in
transient dynamics and cell-cell interaction.  In a long time behavior, 
extinction of multiple cells is repeated, which
leads to itinerancy over successive quasi-stable multi-cellular 
states consisting of different types of cells.
By defining the plasticity of a cellular state, 
it is shown that the plasticity of cells decreases before the
large extinction, from which diversity and plasticity are recovered.
After this switching, decrease of plasticity again occurs, 
leading to the next extinction of multiple
cells.  This cycle is repeated.
Relevance of our results to the development, evolution is
briefly discussed.
\end{abstract}
\section{Introduction}
In multi-cellular organisms, developmental 
process from a single cell or few homogeneous cells with multi-potency 
leads to an organism that consists of various cell types. 
Furthermore, in some multi-cellular organisms such as insects, 
the developmental process is generally accompanied by metamorphosis. 
There, several multi-cellular states are realized, which form 
quasi-stable distribution of cell types.
In the event of metamorphosis,
both the number of cells and the number of cell types decrease drastically, 
and then a different multi-cellular state
is realized.
In general, to consider the diversity of tissues in a multi-cellular
organism, it is important to study how several stable distributions of
different cell types are formed, sustained, or collapsed.
What mechanism causes the transition between different
quasi-stable states with several cell types?
How is it related with the plasticity of cellular states?

As a theoretical framework for developmental process of 
multi-cellular organisms, ``isologous diversification theory'' is proposed
by Kaneko, Yomo, and Furusawa\cite{Kaneko&Yomo2,Kaneko&Yomo3,Furusawa1,Furusawa2}.
In these studies, dynamical systems approach is adopted, to
show that robust developmental process with various cell types emerges 
based on the interplay between intra-cellular dynamics and inter-cellular 
interactions without finely tuned mechanisms. 
On the other hand, there are diversity at a tissue level, i.e., 
different stable distributions consisting
of different cell types, in a usual multi-cellular
organism.  Switches among such quasi-stable multi-cellular states are
also important.  So far spontaneous formation of switches between 
several quasi-stable states is not much 
studied \footnote{In ref\cite{Furusawa2}, a proto-type of the life 
cycle of multi-cellular organisms are shown. However, the results are still
preliminary.}. 
In this paper, we study the dynamics to maintain the diversity of cell types 
and the emergence of switches among several quasi-stable multi-cellular 
states. Here all the cells change their states drastically in each switching
accompanied by deaths of multiple cells.
To quantify the switching mechanism, we define plasticity of 
a cellular state, as changeability of the state against external environmental
change. As will be shown, the switching accompanied by multiple cell deaths
is tightly related with the loss of plasticity.

As the development progresses, several cell types with low plasticity 
increase their number, which leads to extinction of multiple cell types,
following the lack of chemical resources. 
Then a drastic change in environment is caused by the extinction,
which leads to a large change of internal states of all the surviving cells.
Then, different cell types with high plasticity are generated.
From this `undifferentiated state' as in
the initial state, a new cell society with a different set of differentiated cell types is formed. This cycle is repeated. 

For the purpose of the present study, we introduce a new modeling framework
for internal chemical reaction dynamics, that is reaction-diffusion system on 
``chemical species space''. It can be regarded as an extended 
version of Boolean network model\cite{Kauffman1}, so that each variable takes a 
continuous value. 
By adopting this modeling, one can correspond the dynamics of chemical
concentrations with the fixed expression of genes, which is a common
representation for each differentiated cell type.  In this sense,
the symbolic representation of cell types and differentiation rules
are discussed from the dynamics, as is casted as the 5th problem 
in the open problem of Artificial Life\cite{Bedau}.
We then discuss both the stability of 
realizing cell states and their diversity.  
We show that the above cycle to increase and decrease diversity and
plasticity generally appears in our dynamical systems model.

In the next section, we describe the details of our model.
After presenting the 
behavior of a single cell state in terms of the number of attractors
in section III, developmental process is introduced in section IV.
The condition to exhibit cell differentiation by cell-cell 
interaction and switching over several quasi-stable cell society
is presented.
In section V, we propose a feedback process leading to this
switching, and confirm the relationship between the switching
and the plasticity of total cells.
Finally, in section VI, we briefly mention the generality of the
present phenomenon and discuss the relevance to the development and evolution.

\section{model}
The basic strategy of the modeling follows the previous works\cite{Kaneko&Yomo2,Kaneko&Yomo3,Furusawa1,Furusawa2}.
Our model consists of the following three parts:
\begin{itemize}
\item intra-cellular chemical reaction network
\item cell-cell interaction
\item cell division and cell death
\end{itemize}
Now we describe each process. In Fig.1, 
we show the schematic representation of our model.

\begin{itemize}
\item intra-cellular reaction network\\
In general, intra-cellular chemical reactions consist of
the reactions both of genes and 
metabolites.  Genes are set to
be on or off through the interactions between genes and metabolites.
Genes and metabolites both compose reaction networks, while
the time scale of reactions among genes are relatively slower
than those of 
metabolites.
Intra-cellular chemical reaction dynamics of  
our model is simply chosen and abstract so that it satisfies basic features 
described above.

First, we assume a reaction network consisting of many 
products and resource chemicals similarly as that by genes and metabolites.
The concentration of the $j$-th resources in $i$-th cell
is denoted by $u_i^j(t)$ , while that of products is denoted by
and $v_i^j(t)$.
Here,  product chemicals are synthesized autocatalytically by consuming 
corresponding resource chemicals. 
We adopt a variant of Gray-Scott model\cite{Gray&Scott,Pearson} as 
this autocatalytic reaction scheme.
\footnote{ Gray-Scott model is a reaction-diffusion system composed of two chemicals, resource and product. It is a simplified version of 
autocatalytic Selkov 
model that explains self-sustained oscillation of glycolysis.}.
In this paper, we mainly present the results of the case where the number of 
members per each chemical is commonly set to $P=30$.

Next we also assume species-exchanging reactions in each of product and
resource chemicals,
which form two random reaction networks. Each chemical is converted to other
chemicals specified in reaction networks in proportion to the concentration
difference between them.
Here we adopt the same network for products and resources for simplicity, 
though it is plausible that 
these two networks are different from each other.
It is noticed, however, that we also investigated models with
different networks and obtained qualitatively the same results.
The reaction network is represented by a reaction matrix $W{(i,j)}$, 
which takes 1 if there is a reaction from chemical $i$ to 
chemical $j$, and 0 otherwise. Here the reaction network is assumed to be 
symmetrical by considering reversible reaction, that is, 
if $W{(i,j)}$ takes 1, then $W{(j,i)}$ 
also takes 1. This network matrix is chosen randomly 
under the constraint of this symmetry
while the number of reaction paths per each chemical is kept constant. 
The rate constant of species-exchanging reaction is set to $C$ in common 
with all resources, and $D$ in common with all products.
Here we assume that $C$ is larger than $D$, by considering
the difference in the time scales between genes and metabolites. 
These values are fixed throughout the simulation. 
Then, it is necessary to note that the system of random reaction network 
of species-exchanging reactions mentioned above is regarded as 
``diffusion'' on random `chemical species space', and with the reaction
between products and resources, the intra-cellular reaction
is regarded as a reaction-diffusion system. 
In that representation, each cell state corresponds to a discrete
genetic expression pattern.
We can realize multi-attractor states through bifurcations caused by the 
change of parameter $C$ or $D$. 

Since this model consists of the interacting gene network where all the gene 
expressions take continuous values, it includes both the concepts of Boolean 
network model\cite{Kauffman1} and reaction-diffusion system\cite{Turing}.
Thus, we can discuss both stability of each cell state and 
diversity of cell states.
 
\item cell-cell interaction\\
Assuming that environmental medium is completely stirred, we can neglect 
spatial variation of chemical concentrations in it, so that all the cells
share the spatially homogeneous environment.
Here we consider only the diffusion of some chemicals 
through the medium as a minimal form of interaction.
In this model, we assume that only resource chemicals are transported through 
the membrane, in proportion to the concentration difference between the inside 
and the outside of a cell. All the resource chemicals have the same diffusion
coefficient $D_{u}$. 
Each cell grows by taking resource chemicals from the medium and transforms 
them to product chemicals.
$U^j$ is the concentration of $j$-th resources in the medium.
Resource chemicals in the medium are consumed by cells, while we assume 
material bath outside of the medium and impose a flow of 
resource chemicals to the medium in proportion to the difference 
between the concentration in the bath and the medium. 
Again, all the resource chemicals have 
the same diffusion coefficient $D_{U}$ in the medium. 
The concentrations of all resources in the material bath, $U$ are
set to be 1. The parameter $Vol$ is the volume ratio of a medium to a cell,
and $N$ is the number of cells.

\item cell division and cell death\\
Each cell gets resource chemicals from the medium and grows by changing them 
to the product chemicals. 
%Here we assume that the cell volume is proportional to the total amount of product chemicals. 
When the total amount of product chemicals in a cell becomes twice 
the original, then the
cell is assumed to divide, while if it is less than half the original,
then the cell is put to death.
In real biological system, cell division occurs after replication of
DNA (which has smaller diffusion coefficient and cannot penetrate through
the membrane).
Hence these assumptions are rather natural.  
After cell division, each cell volume is set to be half. 
In cell division process, each cell is assumed to be divided into two 
almost equal cells, with some fluctuations. 
To be concrete, chemical concentration $a$($a$ is representation of $u$, $v$)
is divided into $(1+\eta)a_i^{(j)}$ and $(1-\eta)a_i^{(j)}$ respectively,
where $\eta$ is uniform random number over $[-10^{-3},10^{-3}]$.
These fluctuations can give rise to a small variation among cell states, 
which eventually reaches to cell differentiation.

\end{itemize}
Accordingly, the concentration change of each chemical species is given by

\begin{eqnarray}
du_i^j(t)/dt=D_{u}(U^j(t)-u_i^j(t))-u_i^j(t)v_i^j(t)^{2}
+C\sum_{k=1}^{P}W{(j,k)}(u_i^k(t)-u_i^j(t))\nonumber
\end{eqnarray}

\begin{eqnarray}
dv_i^j(t)/dt=u_i^j(t)v_i^j(t)^{2}-Bv_i^j(t)
+D\sum_{k=1}^{P}W{(j,k)}(v_i^k(t)-v_i^j(t))\nonumber
\end{eqnarray}

\begin{eqnarray}
dU^j(t)/dt=\frac{D_{u}}{Vol} \{\sum_{i=1}^{N}u_i^j(t)-U^j(t))
+D_{U}(U-U^j(t)) \} \nonumber
\end{eqnarray}

In the model introduced above, a single cell has many 
fixed-point attractors, in contrast to the previous studies\cite{Kaneko&Yomo2,Kaneko&Yomo3,Furusawa1,Furusawa2}, where a single cell can take one or 
a few attractors.
The reasons why we adopt the present modeling are as follows. 
First many stable cellular states are realized in our model 
that allow for differentiations to many cell types.
Second, as will be shown, by the instability through
the interplay between transient states of cells 
and cell-cell interaction causes cell differentiations, 
although a single cell state eventually reaches 
fixed-point attractors after the transient time.
Third, it is easy to classify cell types as different
fixed point states, and is natural to relate them with different
cellular  states represented by different gene expressions.

\section{Behavior of a single cell state}

\subsection{Initial conditions and methods}
First we investigate the behavior of a single cell state without cell division
and cell death. In all the simulations, 
we set the parameters $C=2.0$, $D=0.020$, $D_{u}=0.50$, $D_{U}=1.0$, $Vol=3.0$,
$A=0.020$, $B=0.060$. 
These values of A$\sim$D correspond to typical values at which 
the original Gray-Scott model forms a self-replicating spot pattern in 
a one dimensional system. 
For numerical integrations, we used $\Delta t=0.010$ and $\Delta x=1.0$,
while we have confirmed that the numerical results are qualitatively 
unchanged by using smaller values for $\Delta t$.
We choose such initial condition that all the resource chemicals are 0.5 and 
all the product chemicals are randomly distributed in a cell.
With this parameter setting, we get fixed point attractor for all cases.

\subsection{Multiple attractors}
Now we investigate how the number of attractor depends on the number of 
reaction paths per each chemical and the number of chemical species. 
The path number is varied from 1 to 6, and the number of chemical species $P$
is varied from 6 to 121.
Examples of fixed point attractors for the cases with
the path number 1(a), 2(b), 4(c), 6(d) 
with the number of chemicals $P=30$ are shown in Fig.2 using a gray scale 
for the concentration of product chemicals. 
All the attractors are fixed points, and can be distinguished easily.
For each number of chemical species, we take one reaction network 
chosen arbitrary, and take 500 different initial conditions to count 
the number of attractors. 
Although we study only one reaction network for each, 
we can get the number dependence rather well, by carrying out
simulations taking various values of $P$.
The dependence of attractor number on $P$ is given in Fig.3.

As the number of species $P$ is increased, the number of attractors 
also increases, while as the number of reaction paths per each chemical 
increases, the number of attractors decreases. 
In case of one reaction path per each chemical, the reaction network 
is decomposed into $P/2$ pairs consisting of 2 chemicals, and 
each pair can take 2 states, so that the 
the number of attractors increases as $2^{P/2}$. 
In case of more reaction paths per each chemical, 
the the number of attractors increases with some power of $P$, that is larger
than 1.

%\subsection{Difference from Boolean network model}
In Boolean network model, the number of attractor also increases with some
power for 2 or 3 reaction paths per each chemical, with an exponent smaller
than 1. Kauffman suggested that this relatively small number of attractors 
against the number of all genes suggests the stability of cell types in living
systems\cite{Kauffman1}. 
We must point out, however, that all the states realized in Boolean network 
model are attractors of a single cell state.
In fact, in real multi-cellular systems, cell states are not necessarily 
determined by possible stable states only of a single cell. 
It is often the case that cells take different states 
realized through cell-cell interaction\cite{Gurdon}.
It is rather plausible that such cellular states are selected that are 
stabilized through mutual relationship between intra-cellular dynamics and 
inter-cellular interaction.  
Hence we study multi-cellular states in the next section.

\section{Developmental process with cell division and cell death}
Now we discuss the change of cellular states under the process of cell 
division and cell death. Here the intra-cellular state and the inter-cellular 
interaction are mutually influenced. By cell-cell interaction, a homogeneous 
cell society of a single cell attractor may be destabilized, so that novel 
states may appear. We study such cell differentiation processes here.

\subsection{Initial conditions and methods}
First, we set an initial cellular state not exactly 
at an attractor of a single cell. 
If we start from an attractor of a single cell state, then the cells 
cannot differentiate by divisions any more. 
With the increase of cell numbers, all the cells compete for taking resources, 
and eventually they go to extinction.
Once the state is on an attractor, the cell division gives two identical cells,
as long as the fluctuation is not large to make a jump to different attractor.
Here we simply adopt the initial condition of a single cell with 
$u_i^j(0)=0.50$ and $v_i^j(0)=0.250+0.01\times rand(j)$, where $j$ is natural 
number over [1,P] and $rand(j)$ is a uniform random variable over [-1,1].

Now we adopt two coarse-grained
representations to analyze the cell differentiation 
events. The first one is introduced in order to distinguish cell types. 
We define a vector representation of each cell state denoted by 
$\overrightarrow{\rm X}$, where each element takes 0 or 1. This value is 
determined by the function given below and the concentration of each product 
chemical. All the cells  that have the same 
$\overrightarrow{\rm X}$ are regarded to belong to the same cell type.
\begin{center}
$\overrightarrow{\rm X}=F(\overrightarrow{\rm V})$,
$\overrightarrow{\rm V}=(v^1,v^2,...,v^{(P)})$,\\
$F(Y)= \cases{
                1 & ($Y \ge 0.5$) \cr
                0 & ($Y < 0.5$)   \cr
}$
\end{center}
Second, to quantify the recursiveness of each cellular state,
we compare the average concentrations of product chemicals in each cell
between two successive division processes. 
$\overrightarrow{\rm V_i(n)}$ denotes the vector representation of the 
average concentration of all product chemicals in $i$-th cell between 
$n-1$th and $n$th cell division. {\overrightarrow{\rm X_i(n)} denotes
a digital representation of {\overrightarrow{\rm V_i(n)},
using the above threshold function $F(Y)$.
As an index for the recursiveness of cellular state, we introduce an inner 
product denoted by $\theta_i(n)$ between {\overrightarrow{\rm X_i(n)} 
and {\overrightarrow{\rm X_i(n-1)}.
If $\theta_i(n)$ takes 1, the cell is regarded to keep the same type 
between successive cell divisions. 
\begin{center}
$\theta_i(n)=\frac{\overrightarrow{\rm X_i(n)} \cdot \overrightarrow{\rm X_i(n-1)}}{|\overrightarrow{\rm X_i(n)}||\overrightarrow{\rm X_i(n-1)}|}$,
$\overrightarrow{\rm X_i(n)}=F(\overrightarrow{\rm V_i(n)})$,\\
$\overrightarrow{\rm V_i(n)}=(\overline{v_i^1},\overline{v_i^2},...,\overline{v_i^{(P)}})$,
$F(Y)= \cases{
                1 & ($Y \ge 0.5$) \cr
                0 & ($Y < 0.5$)   \cr
}$
\end{center}
In this paper, each differentiated cell type is represented as distinctly 
separated chemical states of cells where each cell type can produce 
its own type recursively. 
We study the cell differentiation events with these two representations, 
by which we can distinguish a variety of cell 
states correctly.

\subsection{The condition for cell differentiation}
Here we study the frequency dependence of differentiation event on the 
structure of reaction network. As an index of network structure,
we adopt the number of reaction paths per each chemical.
We investigate four cases where the number of reaction paths per each chemical 
is 1(case (i)), 2(case (ii)), 3(case (iii)), 4(case (iv)) respectively. 
In each case, we take 100 different reaction networks, 
by taking a given initial condition.
We carry out a simulation up to $t=20000$, to check the number of cells
that exist exist at the moment, and the number of cell types. 
Here we note that if cell differentiation does not occur at all, 
all the cells go to extinction by the lack of resource chemicals.
The other parameters and an initial condition are kept at the same values 
as previous section. The frequency of differentiation event 
for each case thus obtained is  given in Fig.4. 

Differentiation event occurs with highest probability in the case (ii). 
This is explained as follows:\\
In the case (i), transient time is shorter and the number of attractors is 
larger than the case (ii) in general. 
In each cell, all the chemicals cannot have global correlation among them 
but differentiate `selfishly' by each cell, so that cooperative relations 
among cells to get resources effectively cannot be formed at a level of cell 
ensemble. As a result, the probability of extinction of all cells is
high. 
On the other hand, in the case (iii) and (iv), transient time is shorter and 
the number of attractors is smaller than the case (ii) in general.
In each cell, all the chemicals have rather strong global correlation among 
them. Hence it is difficult to amplify the small difference to reach cell 
differentiation, so that cooperative relations among cells to get resources 
cannot be well organized. As a result, the probability of extinction of all cells
is again rather high. 
In the case of 5 or more reaction paths per each chemical, this tendency 
is observed more clearly.
In case (ii), however, transient time is longest and the number of attractors 
is the second largest. In each cell, all the chemicals can have 
global correlation among them, and also can form cooperative relations among 
cells to get resources, so that cells can exist with maintaining cell 
differentiation for the longest period. 
Thus, differentiation events occur most probably at this
intermediate number of paths.

\subsection{Long time behavior of coexisting cell group}
Next we investigate the long time behavior of cell differentiation process.
We adopt the case (ii), where the differentiation event is realized most 
clearly. We show an example of temporal evolution that is typically observed.
Existing cell types and its population at every 1000 time are plotted in 
Fig.5. A multi-cellular cooperative state consisting of several cell types 
is formed in the beginning,
and is maintained over a long period, which is much longer than typical 
transient time of single cell state. Then a sudden crisis occurs, and 
after some generations a new multi-cellular cooperative state with 
different cell types is recovered.
This crisis is accompanied by the decrease of the cell number and 
diversity of cell types.
This switching occurs generally, independent of the choice of
the reaction networks.

\section{Further Analysis of the switching process}
\subsection{methods for quantify the plasticity of cell types}
Now we investigate the switching among several
quasi-stable multi-cellular states in more details. 
Since the over all cell states are changed drastically in each switching,
it is an effective way to study the process in terms of the temporal 
change of plasticity of cell types.
Here we define the plasticity of each cell type as changeability 
of the cell state against environmental fluctuations, which are the change 
of concentrations of resource chemicals in the
environment caused by cell divisions and deaths. 
To characterize the plasticity of cell state, we introduce the
following quantity:\\
First we take $v_i^j(0)$, the concentration of $j$-th product chemical 
of the type $i$ 
cell, while we define $v_i^j(f)$ as the concentration of $j$-th product chemical 
of the attractor of the cell type $i$ when
the assumptions of cell division and cell death are eliminated,
to check the attractor state at the fixed environment. 
Then we define the ``attractor distance'' $Dist_i$ of the type $i$ cell, 
as the Euclid distance between $v_i^j(0)$ and $v_i^j(f)$, namely,
$Dist_i=\sqrt{\sum_{j=1}^{P}{(v_i^j(f)-v_i^j(0))^{2}}}\nonumber$.\\
As the distance is smaller, the cell type is closer to a specific attractor.
Since the plasticity of a cell type is the changeability
of the state by the environmental change, the plasticity is 
smaller as the state is  close to a given attractor.
We conjecture that the cell plasticity is characterized
by the attractor distance, and 
the differentiation ability decreases as the attractor distance 
is smaller.

To confirm this conjecture, we have checked the following
two relationships.
First, the relation between the distance and the frequency
of the occurrence of 
re-differentiation event when the a group of cells that
consist only of the single cell type 
is chosen and is put to a new environment. 
The second is the relation between 
the distance and the degree of temporal fluctuations of each cell state when
several cell types co-exist stably for a long term. 
If there are positive correlations in both cases, then the above conjecture 
is confirmed.  These two relations are
given in Fig.6 and Fig.7, for the data shown in the temporal evolution 
in Fig.5.

In Fig.6, we computed
the attractor distances of all cell types that appeared in the
temporal evolution, and divide the whole range of these attractor distances
by 0.1 bin size, and measured the ratio of the occurrence of re-differentiation 
event for all the cell types belonging to each bin.
(Here the population of each cell type is set to be 80. 
If the number is much larger, the homogeneous cell ensemble
grown from the group goes to extinction by the lack of the resources. 
On the contrary, if the number is much smaller, then the effect to internal 
cell states caused by the change of the environment is so large
that the difference of cell types cannot be well distinguished.
Hence we choose this medium number for cell population.)
The result roughly shows sigmoidal function dependence, 
that is, there is a threshold for the attractor distance below which
the corresponding 
cell type loses the ability of re-differentiation drastically. 
This result also supports the initial condition dependence 
of cell differentiation event mentioned previously.

In Fig.7, we computed the temporal fluctuation of each cell type for two
ranges in the temporal evolution, namely, $t=150000\sim250000$ and 
$t=380000\sim480000$, where several cell types co-exist stably. 
We measured the temporal fluctuation of each cell type as the average of all 
the Euclid distances of chemical concentrations
between at the starting time and at some time span later.
Here we take the time span as $t=1000$, so that
in both cases, temporal fluctuations
are calculated from $100$ data samples.  
The result  shows that the temporal fluctuations of the 
chemical concentrations of a cell type are smaller as
the attractor distance gets smaller. 

Hence, it is confirmed that the attractor distance introduced above is valid
as a measure of plasticity of cell type.

\subsection{Mechanism for switching through extinction of many cells}
Now we discuss the switching with multiple cell deaths in
relationship with the loss of plasticity.
Our results are summarized by the following scenario:
At each stage of given quasi-stable multicellular states,
cell types with different degree of plasticity coexist.
Then at each stage,
cell types with relatively high plasticity (i.e., larger
attractor distance) differentiate
to other cell types with lower plasticity, so that the ratio of cell
types with lower plasticity increase gradually.
Then the sustained relation among cell types
to cooperatively use chemical resource chemicals is destroyed, and
extinction of many cell types is resulted.
With this multiple deaths, drastic change of the composition of 
environmental resource chemicals is brought about,
which is large enough to make re-differentiation of some of
surviving cells from the low plasticity state
to a new state with high plasticity. 
Then, including with novel cell types with high plasticity,
a novel multi-cellular state is brought about
allowing for cooperative use of resources.
With this drastic change, a spontaneous switch to novel
multi-cellular state is generated.

To verify this scenario, we computed the temporal change of 
the following four quantities in the temporal evolution of Fig.5:
the average attractor distance for total cells (a),
the total diversity of cell types (b), the average recursiveness 
of cell types over all cells (c), and total number of cell (d).
Here, the average attractor distance is obtained
by averaging the 
attractor distance of each existing cell type with the weight 
of the population ratio. 
The average recursiveness is also computed as the average over
cells whose division event occurs in each $t=1000$ period.
They are plotted in Fig.8, while the population size
for each cell type with given attractor distance is
plotted in Fig.9, around the switching regime.

In Fig.8, as the average attractor distance (a) starts to decrease, and
the diversity of cell types is decreased, while the recursiveness
is increased with strong fluctuations, until
the attractor distance and diversity take the plateau with
low values, when almost complete
recursive production is sustained. After slight decrease of the 
first two values then,  they go up to large values, with the
multiple cell deaths and switching.
After recovering the high plasticity and diversity, they
again decrease to smaller values gradually. 
This process is repeated at each switching event.
The data of Fig.9 also show the accumulation of population
to cell types with
low plasticity before the extinction, and recovery of the types
with high plasticity after it.
Although we show only one example here,
the qualitatively 
same behavior is generally observed at each switching in our system. 
Hence,  the scenario mentioned above is quantitatively demonstrated.

\section{Summary and discussion}

In this paper, we have studied a dynamical systems model of developmental 
process, 
by introducing a new framework, namely, 
reaction-diffusion system on `chemical species space' for modeling 
internal chemical reaction networks. 
This modeling bridges between the two approaches: One is 
reaction-diffusion system proposed by Turing\cite{Turing} and the other is 
Boolean network model proposed by Kauffman\cite{Kauffman1}.
By further including the developmental process, states with many differentiated cell types are
obtained, where the differentiation process to increase the
diversity is observed.
The condition to realize cell 
differentiation by cell-cell interaction is obtained.
When the number of reaction paths per each chemical is larger than 2, 
all the chemicals are correlated strongly and the probability of realizing cell 
differentiation is lower. 
To the contrary, when the number of reaction paths per each chemical 
is 1, then all the chemicals cannot correlate enough and the 
probability of realizing cell differentiation is again lower.
Cell differentiation occurs most probably when the number of 
reaction paths per each chemical is 2, that is, in the intermediate region 
above two cases. These results can be related with those
obtained in the NKCS 
model by Kauffman\cite{Kauffman2}.

In the long time behavior, 
we have found the switching over multi-cellular states
that maintain diverse cell types.
In each multi-cellular state, diverse cell types coexist to
use chemical resources cooperatively, while the switching
is characterized by multiple cell deaths arising from the loss
of diversity of cell types, and the destruction of
the cooperative use of the resources.
This switching behavior is first observed
in our model.

Then we propose that this switching behavior is
characterized by the
loss of plasticity of total cells, that is a general consequence of 
our dynamical systems theory.
In each developmental stage, the irreversible
loss of plasticity is a general course of the development, 
i.e., differentiation from
a cell type with relatively high plasticity to that 
with lower plasticity, so that the ratio of cell types with 
lower plasticity  increase gradually.
Then the sustained relationship among cell types for 
cooperative use of resource chemicals is destroyed, 
which can cause multiple cell deaths. Then, drastic change of 
the composition of environmental resource chemicals is resulted,
which allows for the change of 
surviving cell types from fixed states with low plasticity 
to new states with high plasticity. 
As a results,  new multi-cellular state with novel
cell types is generated. This process is repeated.

This give a feedback process in a system with many degrees of 
freedom to maintain the diversity of cell types, as has
also been discussed as
homeochaos\cite{Ikegami&Kaneko,Kaneko&Ikegami}. 
The plasticity of each cell type in our study corresponds to the mutation
rate of each species in the study of homeochaos. 

The existence of several quasi-stable multi-cellular states
is important to consider the origin of tissues in multi-cellular organisms.
In multi-cellular organisms, several tissues coexist that are
represented as a different distribution of different cell types,
consisting of cells with  the same gene set.
The switching of the states will be important to study
how the life cycle of a multi-cellular 
organism is formed.  In future, the search for a rule of transitions
between successive multi-cellular states will be important.
The dynamics of metamorphosis will be discussed along the line.

The present results are also relevant to
study the evolution.
As evolutionary processes, punctuated equilibria\cite{Gould} are
proposed, while extinction events after long quasi-stationary 
are also discussed.  Successive extinction events of many cells may be
extended to study evolution through extinctions.  Here it is interesting to
note that the distribution of the life-time of quasi-stable multi-cellular states
in our model follows the power law distribution.

On the other hand, the inclusion of genetic mutations to our
model is relevant to study speciation process to multiple species.
Indeed, a theory of sympatric 
speciation with using phenotypic plasticity is
recent proposed\cite{Kaneko&Yomo4,Takagi1}.
In the preliminary studies of our model including
genetic mutation process, we have observed sympatric 
speciation process to form several species,
while with this process, the plasticity of each species defined in this paper
decreases.   
Here, the recovery of plasticity 
of species after extinction of many cells is important to
study open-ended evolution.

We would like to thank T.Yomo and C.Furusawa for stimulating discussions. 
This work is supported by Grants-in-Aid for Scientific Research from
the Ministry of Education, Science and Culture of Japan
(11CE2006, Komaba Complex Systems Life Project, and 11837004).

\begin{figure}[ht]
\begin{center}
\includegraphics[width=13cm,height=7cm]{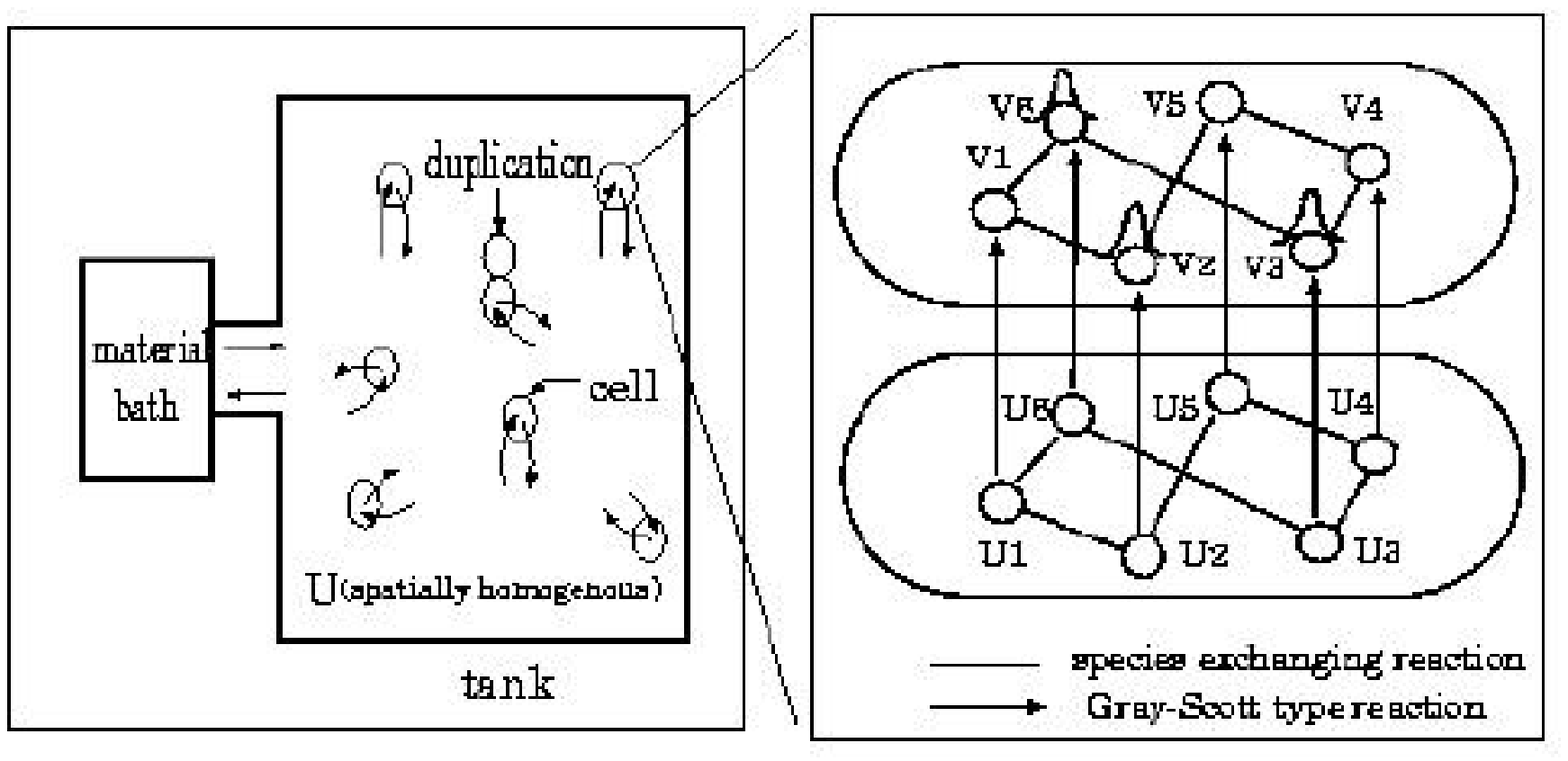}
\vskip 0.25cm
\caption{\small{Schematic representation of our model.}}
\label{fig1}
%\vskip -0.25cm
\end{center}
\end{figure}
\clearpage

\begin{figure}[ht]
\begin{center}
\includegraphics[width=4.5cm,height=2cm]{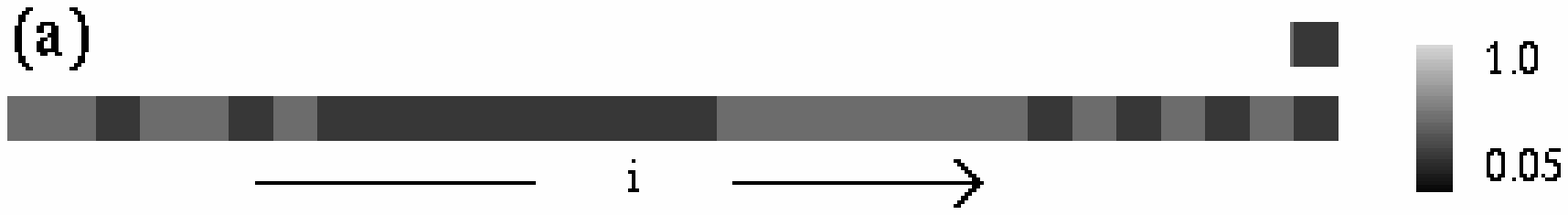}
\includegraphics[width=4.5cm,height=2cm]{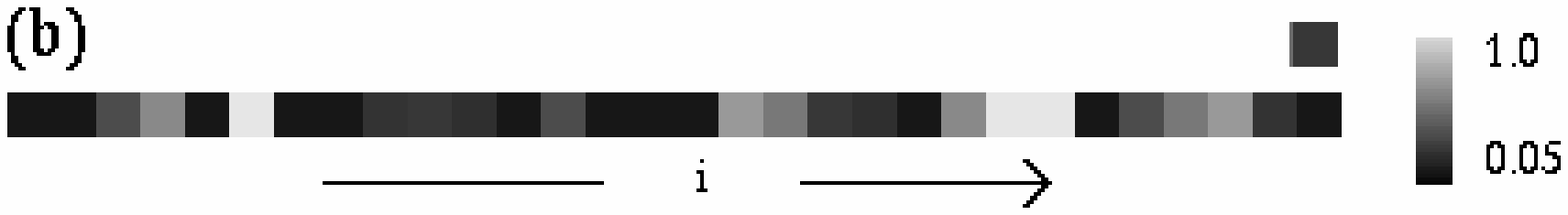}\\
\includegraphics[width=4.5cm,height=2cm]{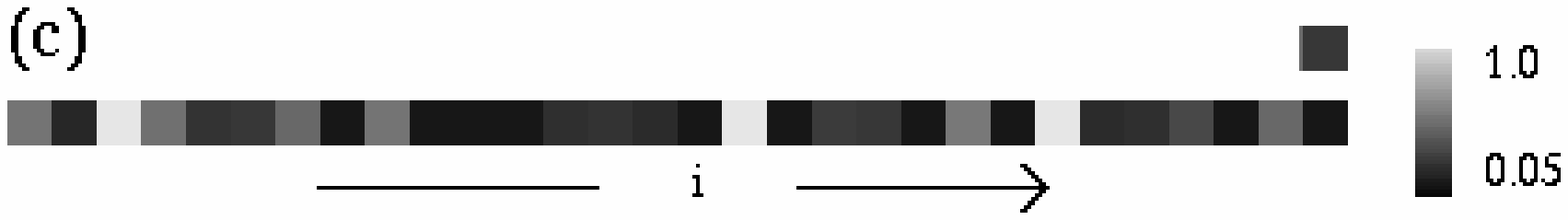}
\includegraphics[width=4.5cm,height=2cm]{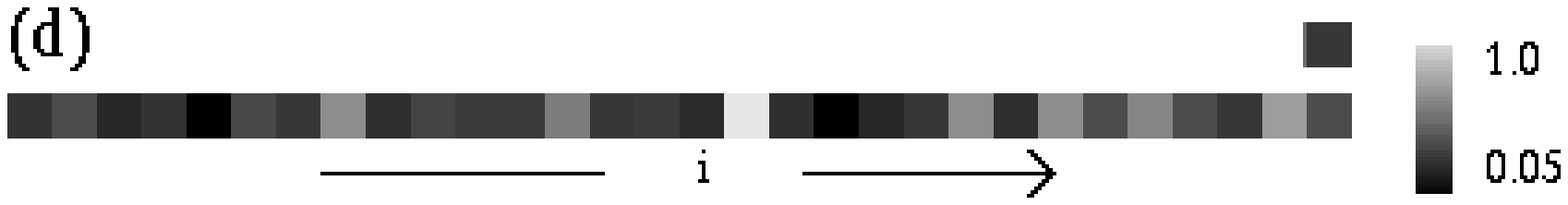}
\vskip 0.25cm
\caption{
\small{Examples of fixed point attractors. The concentrations of all the 
product chemicals for $i=1,2,\cdots,P=30$ are plotted with a gray scale.
%, while the unit size of the figure is also putted. 
As the fixed point attractors, we adopt the state at $t=10000$.
Four cases are studied, namely, one reaction path (a), two reaction paths (b), 
four reaction paths (c), and six reaction paths (d) per each chemical. 
Only one example is shown per each case.  
%Here, the numbers of resource and product chemicals are 30 respectively.
}}
\label{fig2}
\vskip -0.25cm
\end{center}
\end{figure}

\begin{figure}[ht]
\begin{center}
\includegraphics[width=10cm,height=6cm]{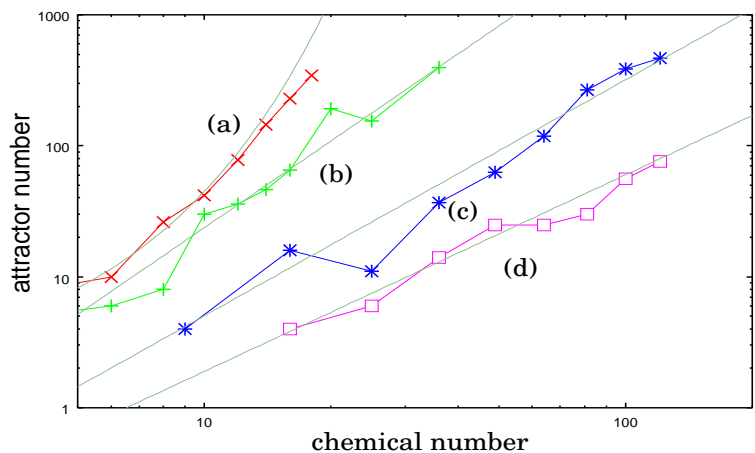}
\vskip 0.25cm
\caption{
\small{Dependence of the number of attractors on the number of chemicals $P$.
Four cases are studied, namely, one reaction path (a), two reaction paths (b), 
four reaction paths (c), and six reaction paths (d) per each chemical. 
For each point of the figure, we take 500 different initial conditions with
the use of only one reaction network. 
Numerically we check the state at $t=10000$ and classify the fixed-point
attractors with the precision of $10^{-4}$.}}
\label{fig3}
%\vskip -0.25cm
\end{center}
\end{figure}

\begin{figure}[ht]
\begin{center}
\includegraphics[width=10cm,height=6cm]{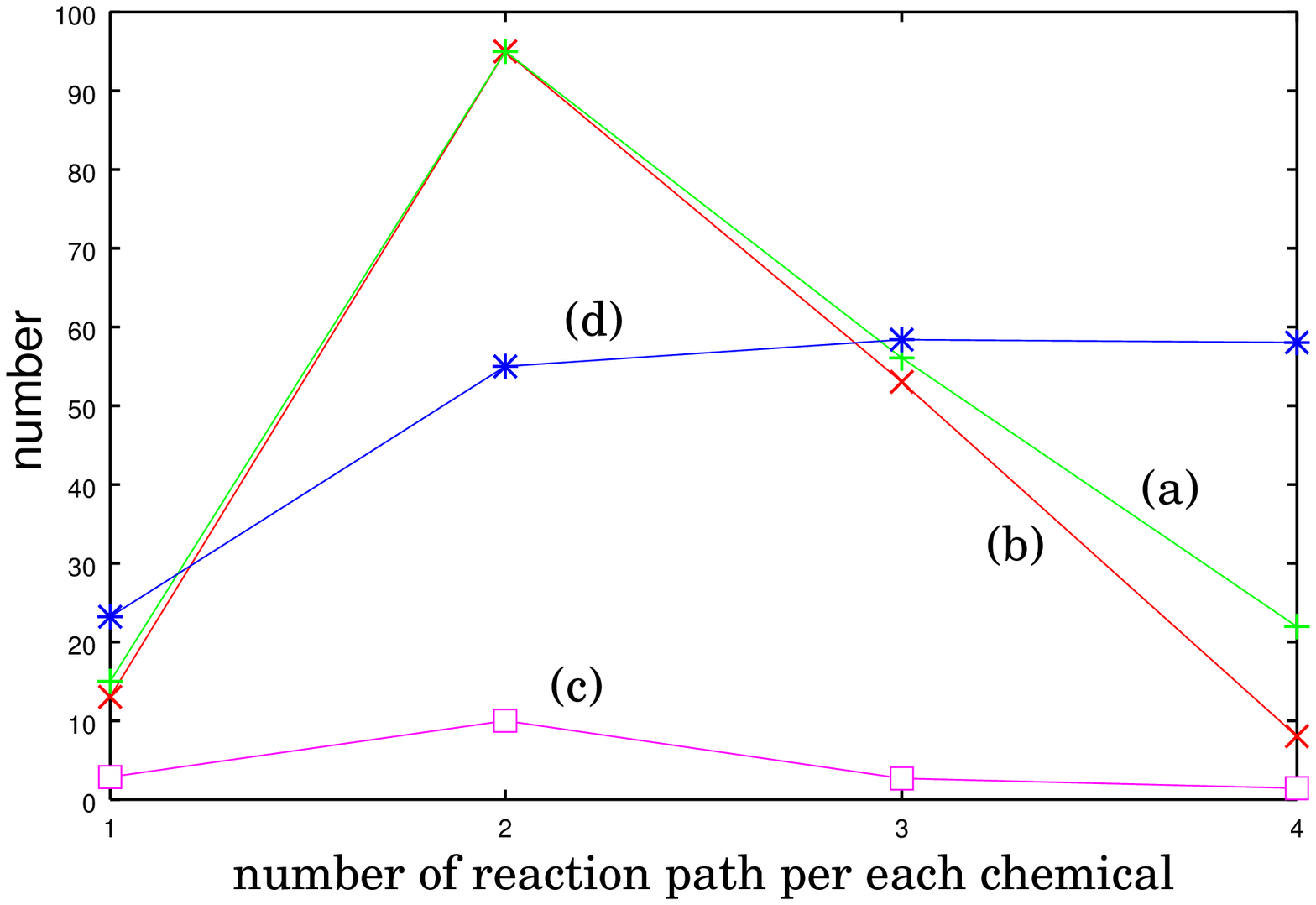}
\vskip 0.25cm
\caption{
\small{The frequency of cell differentiation plotted as a function
of the number of the reaction paths per each chemical. 
We study the cases where the number of reaction paths per each chemical 
is ranged from 1 to 4. 
For each point of the figure, we take 100 different reaction networks and
carry out a simulation up to $t=20000$ with the same initial condition and 
parameters mentioned in the text (without reaction network). 
Accordingly we plot the number of the event where cells exist without 
the whole extinction (a), the number of the event where cell differentiation 
occurs (b), the average number of cell types in the case with cell 
differentiation (c), and the average number of total cells in the case 
without the whole extinction (d).}}
\label{fig4}
%\vskip -0.25cm
%\end{center}
%\end{figure}
%
%\begin{figure}[ht]
%\begin{center}
\includegraphics[width=13cm,height=7cm]{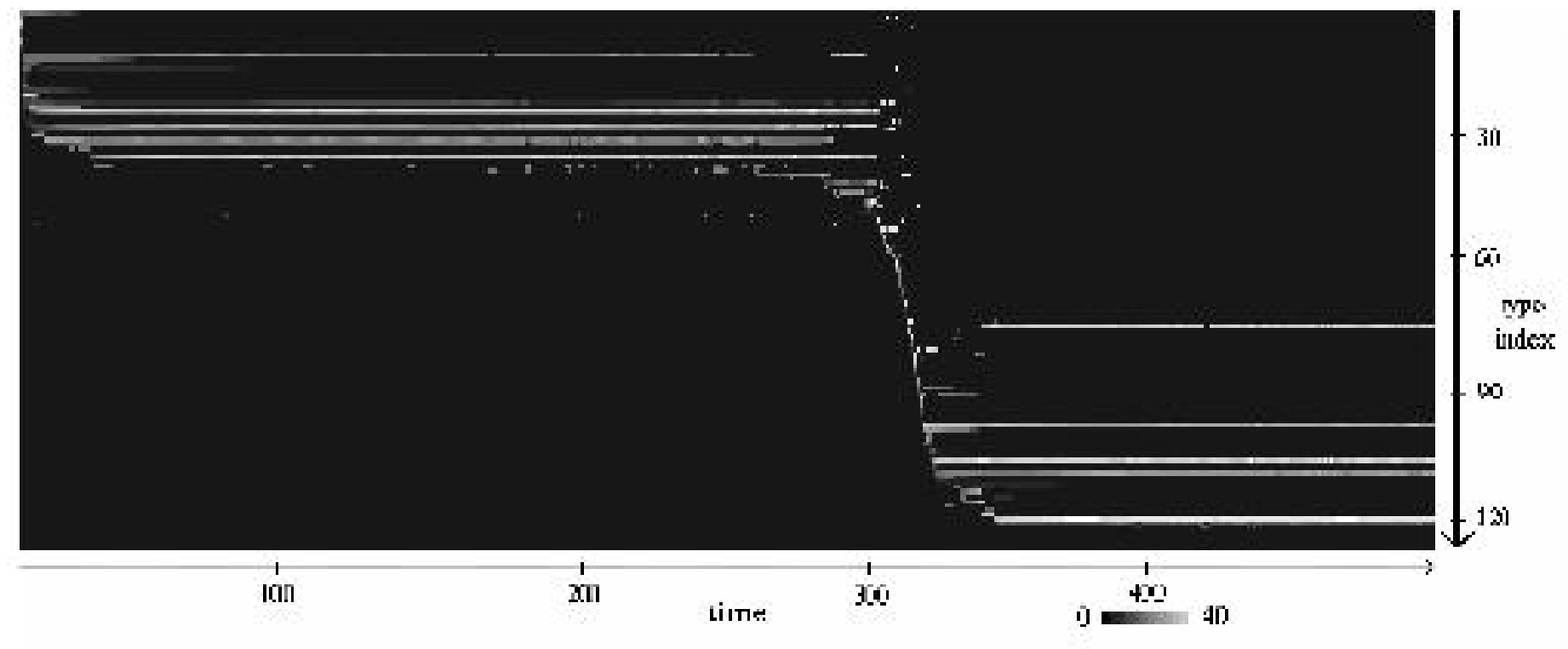}
\vskip 0.25cm
\caption{
\small{An example of the long time behavior of cell differentiation process.
Simulation is carried out up to $t=500000$ starting from a single cell, 
while the data for the total 
cells are sampled by every $1000$ time, to classify all cell types 
and to get the temporal 
change of population of each cell type. 
The initial condition and method for the classification of cell types are 
mentioned in the text.
All cell types are shown in order of their appearance;
a cell type that appears earlier in the simulation has a smaller
index for the cell type.  The
population of each cell type is represented with a gray scale.
Here the unit time of the figures is 1000.}}
\label{fig5}
%\vskip -0.25cm
\end{center}
\end{figure}

\begin{figure}[ht]
\begin{center}
\includegraphics[width=10cm,height=5cm]{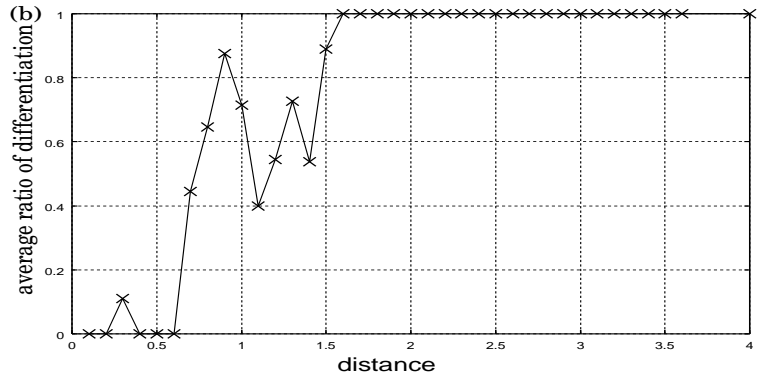}
\vskip 0.25cm
\caption{
\small{The relation between the attractor distance and the possibility of 
the re-differentiation.  We compute attractor distance of all the cell types 
appeared in the temporal evolution shown in Fig.5.
We take an ensemble of cells whose members have the same 
initial condition, and each cell is put it to a new environment and 
see whether cell differentiation occurs or not.
By sampling the data for the attractor distance by 0.1 bin size,
we compute the average ratio of the occurrence of re-differentiation event 
for each bin, to get the relationship between the
differentiation ratio and the attractor distance.
}}\label{fig6}
\vskip -0.25cm
\end{center}
\end{figure}

\begin{figure}[ht]
\begin{center}
\includegraphics[width=10cm,height=5cm]{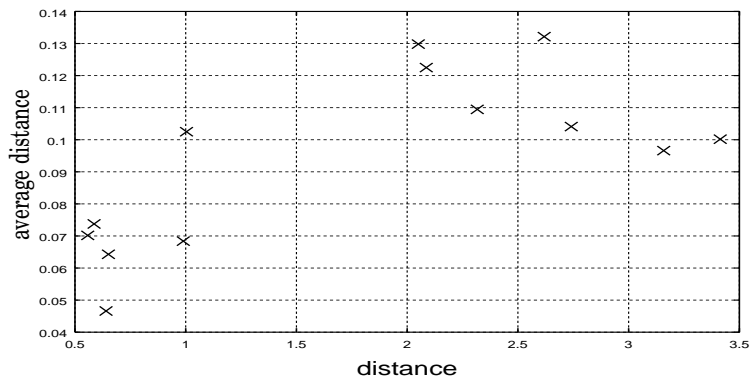}
\vskip 0.25cm
\caption{
\small{Relation between the attractor distance and degree of temporal 
fluctuations of cell states. We calculate a temporal fluctuation of each cell 
type in two ranges in the temporal evolution shown in Fig.5, namely, 
$t=150000\sim250000$ and $t=380000\sim480000$, where several cell types 
co-exist stably. 
Temporal fluctuation of each cell type is computed as the average of all 
the Euclid distances between the starting state and the others.
Here we take the data per $t=1000$.}}
\label{fig7}
%\vskip -0.25cm
\end{center}
\end{figure}

\begin{figure}[ht]
\begin{center}
\includegraphics[width=10cm,height=4.2cm]{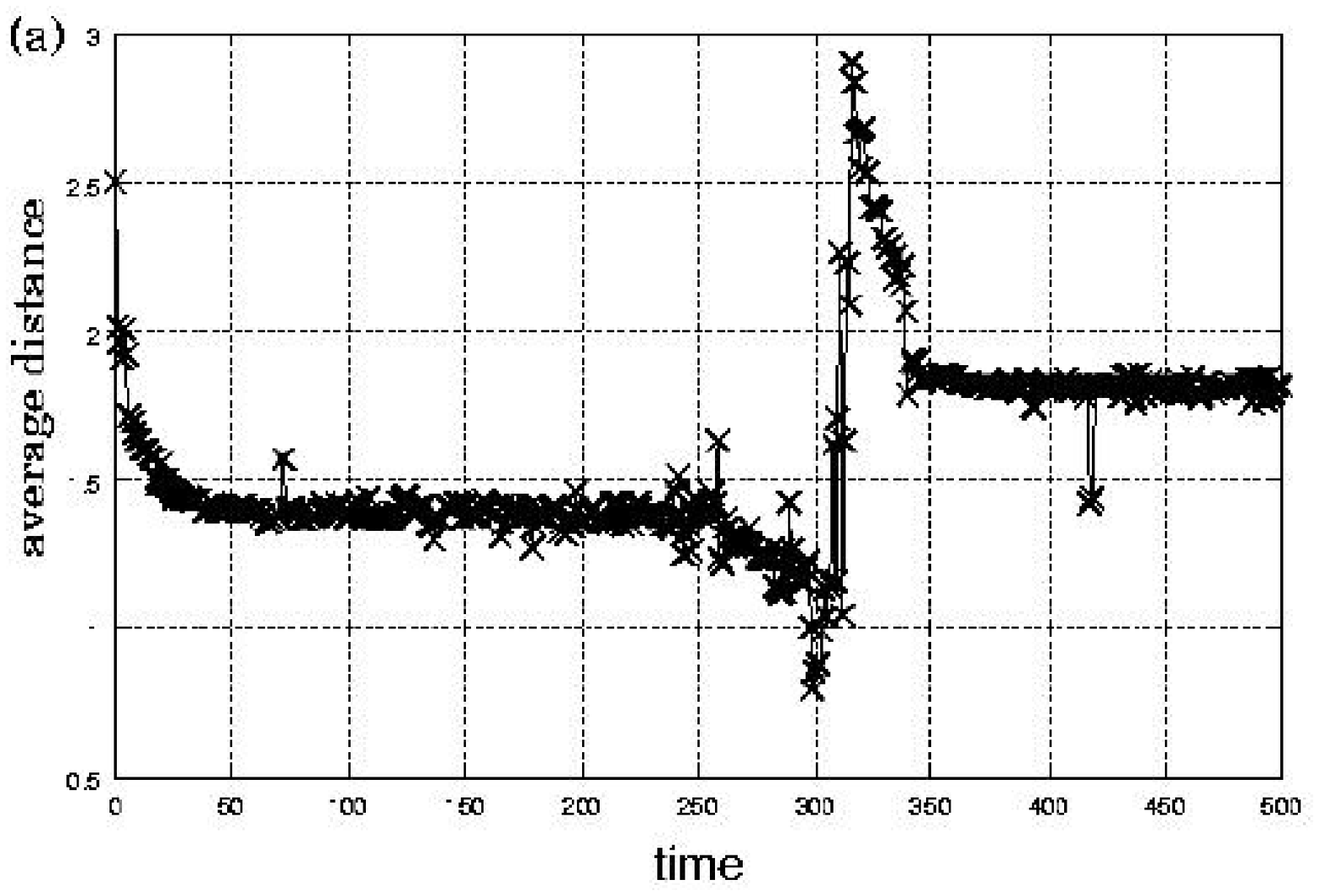}
\includegraphics[width=10cm,height=4.2cm]{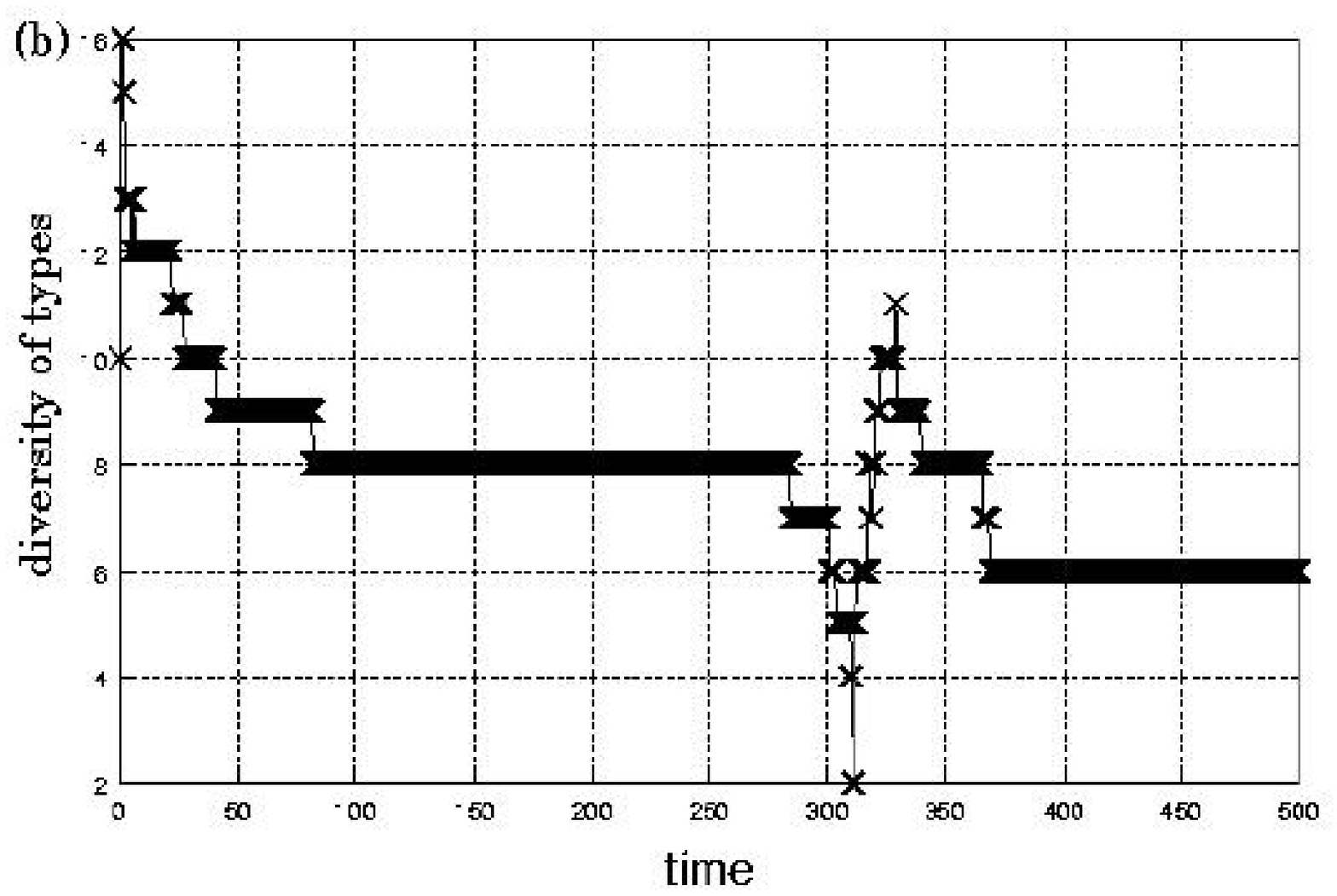}
\includegraphics[width=10cm,height=4.2cm]{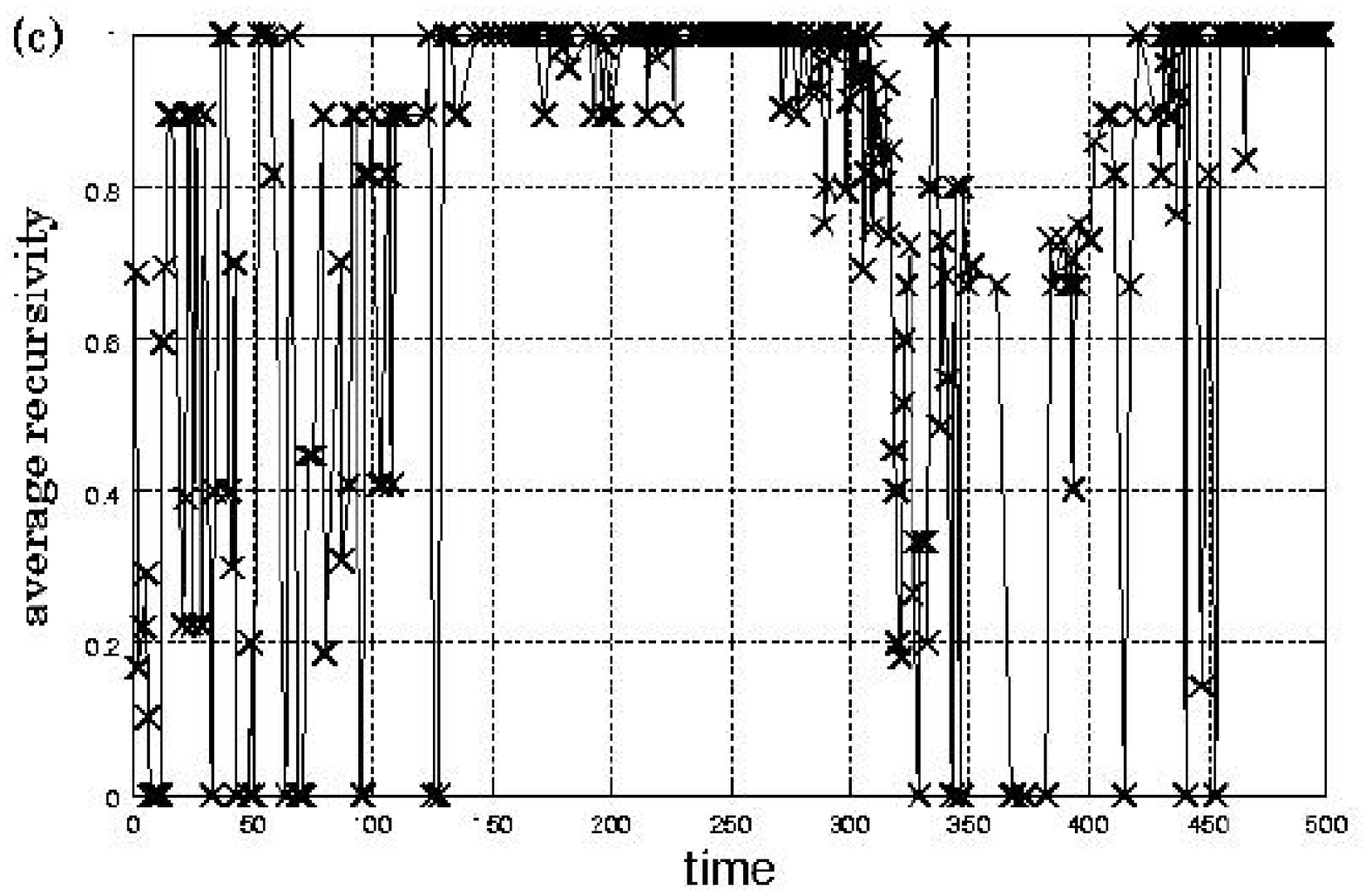}
\includegraphics[width=10cm,height=4.2cm]{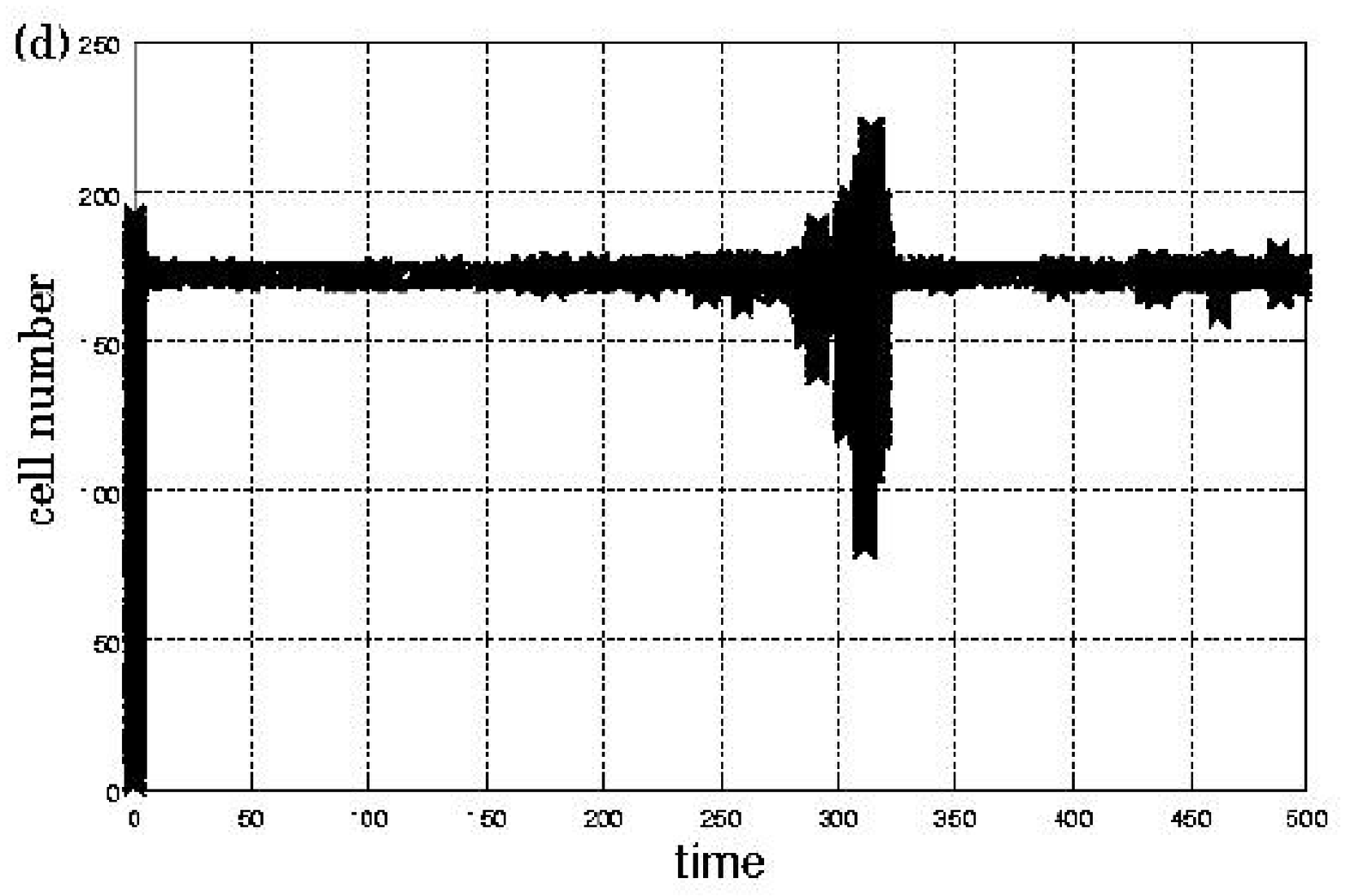}
\caption{
\small{(a):The temporal change of the average attractor distance.
(b):The temporal change of the average diversity over all cells.
(c):The temporal change of the average recursiveness over cell types.
(d):The temporal change of the total cell number.
We computed all the quantities for the data given
in the temporal evolution of Fig.5.
Here the unit time of the figures is 1000.}}
\label{fig8}
\end{center}
\end{figure}
\begin{figure}[ht]
\begin{center}
\includegraphics[width=12cm,height=7cm]{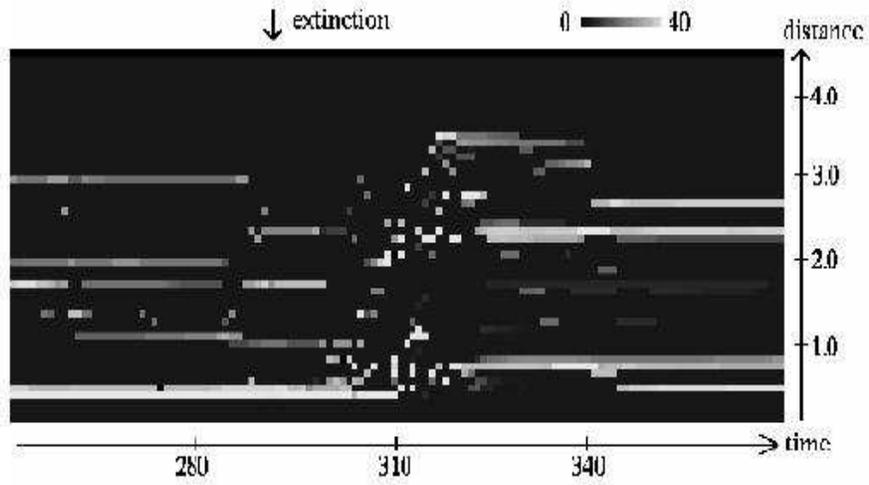}
\vskip 0.25cm
\caption{
\small{The temporal change of the population of existing cell types
around the switching shown in Fig.5.
The attractor distances of the cell types are plotted with their population.
The time where the first extinction of many cells occurs is shown by the arrow.
Here the unit time of the figures is 1000.}}
\label{fig9}
%\vskip -0.25cm
\end{center}
\end{figure}

\begin{figure}[ht]
\begin{center}
\includegraphics[width=12cm,height=7cm]{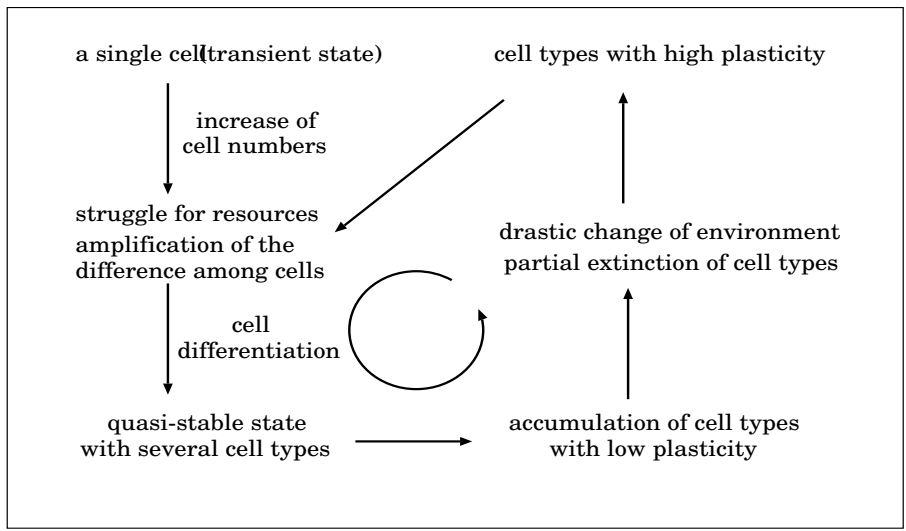}
\vskip 0.25cm
\caption{
\small{Schematic representation of our scenario for the mechanism
to keep the diversity of cell types.}}
\label{fig10}
%\vskip -0.25cm
\end{center}
\end{figure}

\end{document}